\documentclass[12pt,a4paper]{article}
\usepackage{graphicx}
\usepackage[T1]{fontenc}
\usepackage[utf8]{inputenc}
\usepackage{textcomp}
\usepackage[sc,osf]{mathpazo}
\usepackage{a4wide}  
\usepackage{latexsym,amsthm,amsfonts,amsmath,mathrsfs,amssymb}
\usepackage{dsfont}
\usepackage{accents}
\usepackage[nosort]{cite}
\usepackage{booktabs} 
\usepackage[unicode,implicit]{hyperref}
\hypersetup{%
  pdftitle    = {Komar integral and Smarr formula for axion-dilaton black
    holes versus S~duality}
  pdfkeywords = {Black holes, thermodynamics, Noether charge, Wald entropy,
    Komar integral, Smarr formula, first law},
  pdfauthor   = {Dimitrios Mitsios, Tom\'as Ort\'{\i}n and David Pere\~n\'{\i}guez},
  plainpages  = true,
  colorlinks  = true,
  citecolor   = blue,
  urlcolor    = red,
  linkcolor   = black
}
\newcommand{\hepth}[1]{{\tt
\href{http://www.arXiv.org/abs/hep-th/#1}{hep-th/#1}}}
\newcommand{\grqc}[1]{{\tt
\href{http://www.arXiv.org/abs/gr-qc/#1}{gr-qc/#1}}}

\newcommand{\arxiv}[1]{{\tt arXiv:\href{http://www.arXiv.org/abs/#1}{#1}}}
\newcommand{\doi}[1]{{\tt DOI:\href{http://dx.doi.org/#1}{#1}}}

\makeatletter
\@addtoreset{equation}{section}
\makeatother

\begin{document}

\begin{flushright}
\small
IFT-UAM/CSIC-21-065\\
June 15\textsuperscript{th}, 2021\\
\normalsize
\end{flushright}

\vspace{1cm}

\begin{center}

  {\Large {\bf Komar integral and Smarr formula\\[.5cm]
      for axion-dilaton black holes\\[.5cm]
      versus S~duality}}

\vspace{1.5cm}

\renewcommand{\thefootnote}{\alph{footnote}}

{\sl\large Dimitrios Mitsios,}\footnote{Email: {\tt  di.mitsios[at]gmail.com}}
{\sl\large Tom\'{a}s Ort\'{\i}n}\footnote{Email: {\tt tomas.ortin[at]csic.es}}
{\sl\large and David Pere\~n\'{\i}guez,}\footnote{Email: {\tt david.perenniguez[at]uam.es}}

\setcounter{footnote}{0}
\renewcommand{\thefootnote}{\arabic{footnote}}

\vspace{1cm}

{\it Instituto de F\'{\i}sica Te\'orica UAM/CSIC\\
C/ Nicol\'as Cabrera, 13--15,  C.U.~Cantoblanco, E-28049 Madrid, Spain}

\setcounter{footnote}{0}
\renewcommand{\thefootnote}{\arabic{footnote}}
\vspace{1cm}

\vspace{1cm}


{\bf Abstract}
\end{center}
\begin{quotation}
  {\small We construct the Komar integral for axion-dilaton gravity
    using Wald's formalism and momentum maps and we use it to derive a
    Smarr relation for stationary axion-dilaton black holes. While the
    Wald-Noether 2-form charge is not invariant under
    SL$(2,\mathbb{R})$ electric-magnetic duality transformations
    because Wald's formalism does not account for magnetic charges and
    potentials, the Komar integral constructed with it turns out to be
    invariant and, in more general theories, it will be fully
    symplectic invariant. We check the Smarr formula obtained with the
    most general family of static axion-dilaton black holes.}
\end{quotation}

\newpage
\pagestyle{plain}



\section*{Introduction}

In Refs.~\cite{Lee:1990nz,Wald:1993nt,Iyer:1994ys} Wald and his
collaborators Lee and Iyer constructed a powerful formalism that could
be used to prove the first law of black-hole mechanics
\cite{Bardeen:1973gs} and, through this proof, to find the entropy
formula for black-hole solutions of any diffeomorphism-invariant
theory. This formalism has been very successful in absence of matter
fields but it was not clear how to use it on their presence. It is
known that, in many cases, these fields give rise to new terms in the
first law, associated to the possible variations of the conserved
charges associated to them.  It was unclear how these terms could
arise in this formalism since it is based in diffeomorphism invariance
alone and, apparently, the gauge symmetries that ensure the
conservation of the charges that occur in the additional terms of the
first law play no r\^ole whatsoever.

As we have discussed in
Refs.~\cite{Elgood:2020svt,Elgood:2020mdx,Elgood:2020nls},
diffeomorphisms and gauge transformations are, actually, closely
related, because gauge fields are not just tensors. This was one of
the main assumptions in the derivation of the well-known Iyer-Wald
prescription for the entropy Refs.~\cite{Iyer:1994ys}. The
transformation of a gauge field under an isometry which leaves
invariant all the fields of a black-hole solution always induces a
gauge transformation, which, when correctly taken into account
\cite{Prabhu:2015vua} (via \textit{covariant Lie derivatives}, for
instance), gives rise to the missing terms in the first law. If one
uses a tetrad formulation, although the Vielbein is not a matter
field, one must properly take into account that it transforms under
local Lorentz transformations as well \cite{Jacobson:2015uqa} using
the Lie-Lorentz covariant derivative (see
Refs.~\cite{Ortin:2002qb,kn:FF,Ortin:2015hya} and references therein).

Still, terms associated to the variations of charges which are not
associated to gauge symmetries, such as magnetic charges, will not
appear in these derivations of the first law based on Wald's
formalism, while they are known to appear in other derivations of the
first law \cite{Breitenlohner:1987dg}. Terms associated to the
variations of the asymptotic values of the scalars (\textit{moduli})
such as those found in Ref.~\cite{Gibbons:1996af} (see, also,
Ref.~\cite{Astefanesei:2018vga}), will not appear, either. This fact
does not invalidate the first law, but it is a limitation to its
applicability since one cannot study the effects of the variations of
the missing charges.

Smarr formulae \cite{Smarr:1972kt} provide another approach to this
problem. They are closely related to the first law: the scaling
arguments of Refs.~\cite{Kastor:2008xb,Kastor:2010gq} show how the
thermodynamical variables (typically, charges) and their conjugate
thermodynamical potentials must occur in the Smarr formula. This
argument explains why there are no terms associated to the moduli in
the first law if one accepts that the black-hole mass does not depend
on them when it is expressed in terms of the entropy and the conserved
charges.\footnote{This fact follows from the independence of the
  entropy on the moduli, which, to the best of our knowledge, has been
  proven for static, extremal, asymptotically-flat black holes only
  \cite{Ferrara:1997tw,Sen:2005iz}.}

If the black holes under consideration have magnetic charges, then
their Smarr formula must contain a term proportional to them and their
associated potentials. 

As explained in Refs.~\cite{Kastor:2008xb,Kastor:2010gq}, Smarr formulae can
be derived from Komar integrals \cite{Komar:1958wp}. In
Ref.~\cite{Liberati:2015xcp} it was shown how to construct Komar integrals in
general theories using Wald's formalism. The integrand contains a surface term
which is the Noether-Wald charge and a volume term proportional to the
on-shell Lagrangian density. As shown in Ref.~\cite{Ortin:2021ade}, the volume
term can always be expressed as a surface term. Since the variation of the
integral the Noether-Wald charge gives the first law without variations of
magnetic charges and since, as we have argued, the Smarr formula must contain
terms with magnetic charges and potentials, it is not clear how and if those
terms are going to appear. Moreover, electric and magnetic terms must occur in
an electric-magnetic symmetric form in the Smarr formula if the equations of
motion of the theory have that property.

In this paper we want to study if and how this electric-magnetic
duality invariance of the Smarr formula arises from a formalism
(Wald's) which is not electric-magnetic symmetric because only the
gauge transformations which imply the conservation of the electric
charges are taken into account. To this order, in
Section~\ref{sec-axion-dilatongravity}, we are going to study the
static black-hole solutions of a 4-dimensional theory whose equations
of motion are invariant under the archetype of electric-magnetic (or
S-) duality group: ``axion-dilaton gravity,'' which is the bosonic
sector of pure, ungauged, $\mathcal{N}=4,d=4$ supergravity
\cite{Cremmer:1977tt}.  The family of solutions that we are going to
study, found in Ref.~\cite{Kallosh:1993yg} is invariant, as a family,
under the SL$(2,\mathbb{Z})$ duality group and the results obtained
should be automatically invariant under that group. These solutions
will be introduced in Section~\ref{sec-staticBHs}. In
Section~\ref{sec-Komar} we will construct the Komar integral as a
surface integral in a manifestly gauge and diffeomorphism-covariant
form using the momentum maps introduced in
Refs.~\cite{Elgood:2020svt,Elgood:2020mdx,Elgood:2020nls}. In
Section~\ref{sec-checking} we will use the Komar integral to
explicitly test the Smarr formula for the static axion-dilaton black
holes under consideration. A general form of the Smarr formula will,
then, be given in Section~\ref{sec-sduality}, where we will discuss
its electric-magnetic SL$(2,\mathbb{R})$ invariance. Finally,
Section~\ref{sec-discussion} contains our conclusions and some
directions for future work.
 
\section{Axion-dilaton gravity}
\label{sec-axion-dilatongravity}

The 4-dimensional model known as ``axion-dilaton gravity'' is nothing
but the bosonic sector of pure, ungauged, $\mathcal{N}=4,d=4$
supergravity \cite{Cremmer:1977tt} and describes two scalars: the
axion $a$ and the dilaton $\phi$ combined into the complex
\textit{axidilaton} field $\lambda \equiv a +i e^{-2\phi}$ (often
denoted by $\tau$) that parametrizes the coset space
SL$(2,\mathbb{R})/$SO$(2)$, and six 1-form fields
$A^{m}=A^{m}{}_{\mu}dx^{\mu}$ with 2-form field strengths

\begin{equation}
F^{m}=dA^{m}\,,  
\end{equation}

\noindent
coupled to gravity, which we will describe through the Vierbein
$e^{a}=e^{a}{}_{\mu}dx^{\mu}$. The number of 1-forms does not play a relevant
r\^ole if it is larger than one, and can be left undetermined although it has
to be set to six if one wants to embed the solutions of the theory into the
Heterotic Superstring (HST) effective action compactified on a T$^{6}$. The model with just two
1-forms can also be viewed as a model of $\mathcal{N}=2,d=4$ supergravity
coupled to a single vector multiplet, and one can use the powerful
solution-generating techniques developed in that class of models to construct
extremal \cite{Meessen:2006tu,Meessen:2011aa} and non-extremal
\cite{Galli:2011fq,Meessen:2011aa} black-hole solutions.

The action of the theory in the conventions of
Ref.~\cite{LozanoTellechea:1999my}\footnote{The only difference with
  the conventions of
  Refs.~\cite{Kallosh:1992ii,Ortin:1992ur,Kallosh:1993yg,Kallosh:1994ba,Bergshoeff:1996gg}
  is that no imaginary units are introduced with the Hodge
  dualization. These conventions are the same used in
  Refs.~\cite{Elgood:2020svt,Elgood:2020mdx,Elgood:2020nls}.}  in
differential-form language is (summation over repeated $m$ indices is
understood)

\begin{equation}
\label{eq:axiondilatongravityaction}
\begin{aligned}
  S
  & =
\frac{1}{16\pi G_{N}^{(4)}}\int 
  \left[ -\star (e^{a}\wedge e^{b}) \wedge R_{ab}
    +2d\phi\wedge \star d\phi
    +\tfrac{1}{2}e^{4\phi}da \wedge \star da
  \right.
  \\
  & \\
  & \hspace{.5cm}
  \left.
    +2e^{-2\phi}F^{m}\wedge \star F^{m}
+2a F^{m}\wedge F^{m}
\right]
\\
& \\
& \equiv
\int \mathbf{L}\,.
\end{aligned}
\end{equation}

We will set $G_{N}^{(4)}=1$ and we will ignore the normalization factor
$(16\pi)^{-1}$ for the time being.

The equations of motion are defined by

\begin{equation}
  \delta S
  =
  \int\left\{
    \mathbf{E}_{a}\wedge \delta e^{a} + \mathbf{E}_{\phi}\delta\phi
    +\mathbf{E}_{(a)}\delta a  +\mathbf{E}_{m}\wedge A^{m}
    +d\mathbf{\Theta}(\varphi,\delta\varphi)
    \right\}\,,
\end{equation}

\noindent
and given by

\begin{subequations}
  \begin{align}
    \label{eq:Ea}
    \mathbf{E}_{a}
    & =
      \imath_{a}\star(e^{b}\wedge e^{c})\wedge R_{bc}
      +2\left(\imath_{a}d\phi \star d\phi +d\phi\wedge \imath_{a}\star d\phi\right)
    \nonumber \\
    & \nonumber \\
    & \hspace{.5cm}
      +\tfrac{1}{2}e^{4\phi}\left(\imath_{a}da \star da +da\wedge
      \imath_{a}\star da\right)
      +2e^{-2\phi}\left(\imath_{a}F^{m}\wedge \star F^{m} -F^{m}\wedge \imath_{a}\star
      F^{m}\right)\,,
    \\
    & \nonumber \\
    \mathbf{E}_{\phi}
    & =
      -4d\star d\phi +2e^{4\phi}da \wedge \star da -4e^{-2\phi}F^{m}\wedge \star F^{m}\,,
    \\
    & \nonumber \\
    \mathbf{E}_{(a)}
    & =
 -d\left(e^{4\phi}\star da\right) +2F^{m}\wedge F^{m}\,,
    \\
    & \nonumber \\
    \mathbf{E}_{m}
    & =
      -4d F_{m}\,,
  \end{align}
\end{subequations}

\noindent
where we have defined the dual 2-form field strength

\begin{equation}
  \label{eq:dualfieldstrengthsdef}
  F_{m} \equiv \tfrac{1}{4}\frac{\delta S}{\delta F^{m}}
  = e^{-2\phi} \star F^{m} +a F^{m}\,.
\end{equation}

Furthermore,

\begin{equation}
  \mathbf{\Theta}(\varphi,\delta\varphi)
  =
  -\star (e^{a}\wedge e^{b})\wedge \delta \omega_{ab} +4\star d\phi\delta\phi
  +e^{4\phi} \star da\delta a +4F_{m}\wedge \delta A^{m}\,.
\end{equation}

Since the Maxwell equations tell us that the $F_{m}$s are closed
on-shell, we can introduce a dual 1-form field $A_{m}$ defined by

\begin{equation}
F_{m}=dA_{m}\,.  
\end{equation}

\section{Static dilaton-axion black hole solutions}
\label{sec-staticBHs}

The most general family of non-extremal, static, black holes with
non-trivial dilaton, axion and electromagnetic fields was obtained in
Ref.~\cite{Kallosh:1993yg}.\footnote{These solutions were obtained by
  an SL$(2,\mathbb{R})$ rotation of those found in
  Ref.~\cite{Ortin:1992ur}. The case with a single 1-form had been
  dealt with in Ref.~\cite{Shapere:1991ta}, but it is qualitatively
  different since these solutions can have electric and magnetic
  charges and vanishing axion. In their turn, the solutions of
  Ref.~\cite{Ortin:1992ur} are a generalization of those in
  Ref.~\cite{Kallosh:1992ii}, which were originally discovered by
  Gibbons and Maeda in Refs.~\cite{Gibbons:1982ih,Gibbons:1987ps}. The
  single-vector case was rediscovered by Garfinkle, Horowitz and
  Strominger in Ref.~\cite{Garfinkle:1990qj} and it is the solution on
  which the SL$(2,\mathbb{R})$ rotation was performed in
  Ref.~\cite{Shapere:1991ta}. Stationary generalizations (inclusion of
  NUT charge) were constructed in \cite{Kallosh:1994ba} and, for the
  extremal case, using supersymmetry and spinorial techniques, in
  Ref.~\cite{Tod:1995jf} (see also Ref.~\cite{Bergshoeff:1996gg}.)
  Finally, the most general, non-extremal, stationary black-hole
  solution of the model was constructed in
  Ref.~\cite{LozanoTellechea:1999my}.} In the notation of
Ref.~\cite{LozanoTellechea:1999my}, these solutions take the
form\footnote{This presentation of the solutions uses only the time
  components of the original and dual vector fields. As we are going
  to see, this information in enough to fully reconstruct all the
  components of these vectors.}

\begin{eqnarray}
ds^{2}                 & = &
e^{2U}dt^{2}-e^{-2U}dr^{2}-R^{2}d\Omega^{2}_{(2)}\,,
\nonumber \\
\nonumber \\
\lambda             & = &
\frac{\lambda_{\infty}r+\lambda^{*}_{\infty}\Upsilon}{r
+\Upsilon}\,,
\nonumber \\
\\
A^{m}{}_{t}        & = &
e^{\phi_{\infty}}R^{-2}[\Gamma^{m}(r+\Upsilon)+\mathrm{c.c.}]\,,
\nonumber \\
\nonumber \\
A_{m\, t} & = &
e^{\phi_{\infty}}R^{-2}[\Gamma^{m}(\lambda_{\infty}r+
            \lambda^{*}_{\infty} \Upsilon)+\mathrm{c.c.}]\,,
            \nonumber 
\end{eqnarray}

\noindent
where the functions that occur in the metric are

\begin{eqnarray}
e^{2U}  & = & R^{-2}(r-r_{+})(r-r_{-})\;,  \qquad
r_{\pm}    = M\pm r_{0}\,,
              \nonumber \\
  & &  \\
R^{2}   & = &  r^{2}-|\Upsilon|^{2}\,,
\hspace{2,8 cm}r_{0}^{2}=M^{2}+|\Upsilon|^{2}
-4\Gamma^{m} \Gamma^{m\,*}\,. \nonumber 
\end{eqnarray}

In these functions, $M$ is the ADM mass, the constants $\Gamma^{m}$ are
related to the complex electromagnetic charges,
$\lambda_{\infty}=a_{\infty}+ie^{-2\phi_{\infty}}$ is the asymptotic value of
the axidilaton and $\Upsilon=\Sigma+i\Delta$ is the axidilaton charge. All
these parameters are defined by the asymptotic expansions

\begin{subequations}
  \begin{align}
    g_{tt}
    & \sim  1-\frac{2M}{r}\,,
    \\
    & \nonumber \\
    \lambda
    & \sim
    \lambda_{\infty}-ie^{-2\phi_{\infty}} \frac{2\Upsilon}{r}\,,
    \\
    & \nonumber \\
    \tfrac{1}{2}\left[F^{m}{}_{tr}+i\star F^{m}{}_{tr}\right]
    & \sim
    \frac{e^{+\phi_{\infty}}\Gamma^{m}}{r^{2}}
    =\frac{e^{+\phi_{\infty}}(Q^{m}+iP^{m})/2}{r^{2}}\,.
  \end{align}
\end{subequations}

\noindent
The axidilaton charge is not an independent parameter. In accordance with the
no-hair theorem, it is a function of the ADM mass and the electric and
magnetic charges

\begin{equation}
  \label{eq:upsilongamma}
  \Upsilon
  =
  -\frac{2}{M}\Gamma^{m\, *}\Gamma^{m\, *}\,.
\end{equation}

The singularity is hidden under a horizon located at $r=r_{+}$ if
$r_{0}^{2}> 0$, and it is hidden or coincides with it (but still is invisible
for external observers) if $r_{0}=0$.

The solution has been expressed, by convenience, using only the electric
components of the 1-forms and the dual 1-forms. The magnetic components can be
obtained as follows. From the definition of the dual 2-form field strengths
Eq.~(\ref{eq:dualfieldstrengthsdef}), we get

\begin{equation}
    F_{m\, rt}
    =
    \frac{e^{-2\phi}}{R^{2}\sin{\theta}}F^{m}{}_{\theta\varphi} +aF^{m}{}_{rt}\,,  
\end{equation}

\noindent
so

\begin{equation}
  F^{m}{}_{\theta\varphi}
 =
  e^{2\phi}R^{2}\sin{\theta}   \left( F_{m\, rt}-aF^{m}{}_{rt}\right)
=
  2e^{\phi_{\infty}}\Im\mathrm{m}(\Gamma^{m})\sin{\theta}\,.
\end{equation}

\noindent
The gauge field $A^{m}$, then, has to be defined in two patches. On
the $z\geq -\epsilon$ patch it is given by the 1-form

\begin{equation}
  A^{m\,+}
  =
  e^{\phi_{\infty}}R^{-2}[\Gamma^{m}(r+\Upsilon)+\mathrm{c.c.}]dt
  +2e^{\phi_{\infty}}\Im\mathrm{m}(\Gamma^{m})(1-\cos{\theta})d\varphi\,,
\end{equation}

\noindent
which is regular in that region\footnote{The Dirac string singularity
  of this 1-form lies in the negative $z$ axis.}  and in the
$z\leq +\epsilon$ patch, it is given by the 1-form

\begin{equation}
  A^{m\,-}
  =
  e^{\phi_{\infty}}R^{-2}[\Gamma^{m}(r+\Upsilon)+\mathrm{c.c.}]dt
  -2e^{\phi_{\infty}}\Im\mathrm{m}(\Gamma^{m})(1+\cos{\theta})d\varphi\,,
\end{equation}

\noindent 
which is also regular in that patch. $A^{m\,+}$ and  $A^{m\,-}$ differ by 
the gauge transformation

\begin{equation}
A^{m\,+}-A^{m\,-}
  =
d\left[  4e^{\phi_{\infty}}\Im\mathrm{m}(\Gamma^{m})\varphi\right]\,.
\end{equation}

We can also compute the complete dual vector fields. From the definition
Eq.~(\ref{eq:dualfieldstrengthsdef}) we find that

\begin{equation}
  F_{m\, \theta\varphi}
  =
  e^{-2\phi}R^{2}\sin{\theta}F^{m}{}_{tr}+aF^{m}{}_{\theta\varphi}
  =
  2e^{\phi_{\infty}}\left\{e^{-2\phi_{\infty}}\Re\mathrm{e}(\Gamma^{m})
    +a_{\infty}\Im\mathrm{m}(\Gamma^{m})\right\}\sin{\theta}\,,
\end{equation}

\noindent
and

\begin{subequations}
  \begin{align}
  A_{m}{}^{+}
  & =
    e^{\phi_{\infty}}R^{-2}[\Gamma^{m}(\lambda_{\infty}r+\lambda_{\infty}^{*}\Upsilon)
    +\mathrm{c.c.}]dt
    \nonumber \\
    & \nonumber \\
    & \hspace{.5cm}
    +2e^{\phi_{\infty}}\left\{e^{-2\phi_{\infty}}\Re\mathrm{e}(\Gamma^{m})
    +a_{\infty}\Im\mathrm{m}(\Gamma^{m})\right\}(1-\cos{\theta})d\varphi\,,
    \\
    & \nonumber \\
  A_{m}{}^{-}
  & =
    e^{\phi_{\infty}}R^{-2}[\Gamma^{m}(\lambda_{\infty}r+\lambda_{\infty}^{*}\Upsilon)
    +\mathrm{c.c.}]dt
    \nonumber \\
    & \nonumber \\
    & \hspace{.5cm}
    -2e^{\phi_{\infty}}\left\{e^{-2\phi_{\infty}}\Re\mathrm{e}(\Gamma^{m})
    +a_{\infty}\Im\mathrm{m}(\Gamma^{m})\right\}(1+\cos{\theta})d\varphi\,,
  \end{align}
\end{subequations}

\noindent
in the same two patches, and

\begin{equation}
A_{m}{}^{+}-A_{m}{}^{-}
  =
d\left\{  4e^{\phi_{\infty}}\left[e^{-2\phi_{\infty}}\Re\mathrm{e}(\Gamma^{m})
    +a_{\infty}\Im\mathrm{m}(\Gamma^{m})\right]\varphi\right\}\,.
\end{equation}

The Hawking temperature and Bekenstein-Hawking entropy of these black holes
are given by

\begin{subequations}
  \begin{align}
    T
    & =
      \frac{1}{4\pi}\partial_{r}g_{tt}(r_{+})
      =
      \frac{r_{0}}{2\pi R^{2}(r_{+})}\,,
    \\
    & \nonumber \\
    S
    & =
      \pi R^{2}(r_{+})\,.
  \end{align}
\end{subequations}

Observe that, as usual in 4-dimensional, static black holes

\begin{equation}
2ST = r_{0}\,.  
\end{equation}

\noindent
Then, it is not difficult to find a Smarr-type relation adding the ADM mass to
the above relation:

\begin{equation}
  \label{eq:Smarrformula}
  \begin{aligned}
    M
    & =
    2ST+M-r_{0}
    =
    2ST+r_{-}
    =
    2ST+\frac{r_{-}r_{+}}{r_{+}}
    =
    2ST+\frac{M^{2}-r_{0}^{2}}{r_{+}}
    \\
    & \\
    &     =
    2ST+\frac{4\Gamma^{m}\Gamma^{m\,*}-|\Upsilon|^{2}}{r_{+}}
    =
    2ST + \frac{Q^{m}}{r_{+}}Q^{m}
      +\frac{P^{m}}{r_{+}}P^{m} -\frac{\Sigma}{r_{+}}\Sigma
    -\frac{\Delta}{r_{+}}\Delta\,.
  \end{aligned}
\end{equation}

This relation is correct (by construction) and, looking at it, it is tempting
to conclude that the $1/r_{+}$ terms (including those associated to the scalar
charges) can immediately be identified with potentials on the
horizon. However, as we are going to see, $r_{-}$ can be rewritten in other
ways in which only potentials associated to the electric and magnetic charges
occur. Note that the usual scaling argument does not allow for terms including
scalar charges or potentials because, by the no-hair theorem, these cannot be
independent. Indeed, the Komar charge leaves only room for electric and
magnetic potentials and charges, and, as we are going to see, the integral
gives the above relation, although in a highly non-trivial way.

\section{Komar integral}
\label{sec-Komar}

As explained, for instance, in Refs.~\cite{Kastor:2008xb,Kastor:2010gq} Smarr
formulae \cite{Smarr:1972kt} can be systematically obtained from Komar
integrals \cite{Komar:1958wp}. These can be constructed using Wald's formalism
following Ref.~\cite{Liberati:2015xcp}, rewriting the volume integral terms as
surface terms as explained in Ref.~\cite{Ortin:2021ade}. In that reference,
though, the integrand of the surface integral was determined after explicit
evaluation of the Lagrangian density on a particular family of solutions and,
here, we are going to show how that integrand can be found in
general.\footnote{It is assumed, though, that we are restricting ourselves to
  solutions admitting a timelike Killing vector with a Killing horizon.}

Let us review the construction of the Komar charge and integral in
Ref.~\cite{Liberati:2015xcp,Ortin:2021ade}.  It is not difficult to see that,
on-shell\footnote{We are going to use the symbol $\dot{=}$ for identities that
  only hold on-shell.} and for a Killing vector $k$ that generates a symmetry
of the whole field configuration

\begin{equation}
\mathbf{J}[k] \dot{=}\imath_{k}\mathbf{L}\,.  
\end{equation}

\noindent
On the other hand, for any vector field $\xi$, we have the off-shell (local)
identity

\begin{equation}
\mathbf{J}[\xi] = d\mathbf{Q}[\xi]\,.
\end{equation}

\noindent
Combining these two relations, we find that, on-shell and for a Killing vector
$k$ that generates a symmetry of the whole field configuration

\begin{equation}
d\mathbf{Q}[k]-\imath_{k}\mathbf{L} \dot{=} 0\,.  
\end{equation}

\noindent
However, if $k$ generates a symmetry of the whole field configuration,

\begin{equation}
0\dot{=}\pounds_{k}\mathbf{L}= d\imath_{k}\mathbf{L}\,,  
\end{equation}

\noindent
which implies the local existence of a $(d-2)$-form $\omega_{k}$ such that

\begin{equation}
d \omega_{k} \dot{=} \imath_{k}\mathbf{L}\,.  
\end{equation}

\noindent
It follows that, under the aforementioned conditions,

\begin{equation}
  \label{eq:Komaridentity}
d\left\{\mathbf{Q}[k]-\omega_{k}\right\}\dot{=}0\,.  
\end{equation}

\noindent
and we can define the Komar integral over the codimension-2 surface $\Sigma^{d-2}$
Ref.~\cite{Ortin:2021ade}

\begin{equation}
  \label{eq:Komarintegral}
  \mathcal{K}(\Sigma^{d-2})
  =
  (-1)^{d-1}\int_{\Sigma^{d-2}}\left\{\mathbf{Q}[k]-\omega_{k}\right\}\,.
\end{equation}

Smarr formulae for black-hole spacetimes are obtained by integrating the
identity Eq.~(\ref{eq:Komaridentity}) on hypersurfaces $\Sigma$ with
boundaries at the horizon and spatial infinity $\partial \Sigma_{h}$ (usually,
the bifurcation surface) and $\partial\Sigma_{\infty}$, respectively upon use
of Stokes theorem:

\begin{equation}
  \mathcal{K}(\partial\Sigma_{\infty})
  =
  \mathcal{K}(\partial\Sigma_{h})\,.
\end{equation}

Using the techniques developed in
Refs.~\cite{Elgood:2020svt,Elgood:2020mdx,Elgood:2020nls} and some of the
results found in them, we can readily find  the Noether-Wald charge for
axion-dilaton gravity:

\begin{equation}
    \mathbf{Q}[\xi]
        =
        \star (e^{a}\wedge e^{b})
        e^{-2\phi}P_{\xi\, ab} -4P^{m}{}_{\xi}F_{m}\,.
\end{equation}

Here,

\begin{equation}
P_{\xi\, ab}  =\nabla_{[a}\xi_{b]}\,.
\end{equation}

\noindent
Also, the functions $P^{m}{}_{\xi}$ can be understood as the parameters of
compensating gauge transformations of the 1-forms with the property that, when
$\xi=k$, they satisfy the relations

\begin{equation}
  \label{eq:electricmomentummapdef}
dP^{m}{}_{k} = -\imath_{k}F^{m}\,, 
\end{equation}

\noindent
that define the \textit{momentum maps} associated to the Killing vector $k$ and
the gauge fields $A^{m}$. Although this is a gauge-invariant definition,
these objects are defined up to an additive constant. Since they can be
interpreted as electrostatic potentials, the constant can be determined by
a sensible boundary condition, such as the vanishing of the potentials at
spatial infinity.

In order to compute $\omega_{k}$, we have to determine the on-shell value of
the Lagrangian density $\mathbf{L}$ first, for a generic solution. In this
case, it is enough to use the trace of the Einstein equations
Eqs.~(\ref{eq:Ea}).  In differential-form language, to take the trace we must
compute $e^{a}\wedge \mathbf{E}_{a}$, taking into account that, for a $p$-form
$\omega^{(p)}$,

\begin{equation}
  e^{a}\wedge \imath_{a} \omega^{(p)}
  = p  \omega^{(p)}\,.
\end{equation}

\noindent
We get

\begin{equation}
    \label{eq:traceEa}
    \begin{aligned}
      e^{a}\wedge \mathbf{E}_{a}
      & =
      -2\left\{-e^{-2\phi} \star (e^{c}\wedge
        e^{d})\wedge R_{cd} +2d\phi\wedge \star d\phi +\tfrac{1}{2}e^{4\phi}
        da\wedge \star da\right\}
      \\
      & \\
      & =
      -2 \left\{\mathbf{L} - 2e^{-2\phi}F^{m}\wedge \star F^{m}-2a F^{m}\wedge F^{m}
\right\}\,,
    \end{aligned}
\end{equation}

\noindent
so

\begin{equation}
  \mathbf{L}
  \dot{=}
  2e^{-2\phi}F^{m}\wedge \star F^{m}+2a F^{m}\wedge F^{m}
  =
  2F^{m}\wedge F_{m}\,,
\end{equation}

\noindent
and

\begin{equation}
 \imath_{k} \mathbf{L}
  \dot{=}
  2\imath_{k}F^{m}\wedge F_{m}+  2F^{m}\wedge \imath_{k} F_{m}\,.
\end{equation}

In order to find $\omega_{k}$ for general configurations, we are going to use
the definition of the (\textit{electric}) momentum maps
Eq.~(\ref{eq:electricmomentummapdef}) but we need to define their magnetic
duals. Since, by assumption, the dual field strengths are left invariant by
the isometry generated by $k$,

\begin{equation}
0=\pounds_{k}F_{m} = d\imath_{k}F_{m} +\imath_{k}dF_{m} \dot{=}  d\imath_{k}F_{m}\,, 
\end{equation}

\noindent
where we have used the Maxwell equations. Then, locally, there are functions
$P_{m\,k}$ (\textit{magnetic momentum maps}) such that 

\begin{equation}
  \label{eq:magneticmomentummapdef}
dP_{m\,k} \dot{=} -\imath_{k}F_{m}\,. 
\end{equation}

Thus, upon use of the Maxwell equations and Bianchi identities,

\begin{equation}
 \imath_{k} \mathbf{L}
  \dot{=}
  -2dP^{m}{}_{k}\wedge F_{m}-2F^{m}\wedge dP_{m\,k}
  \dot{=}
  d\left\{-2P^{m}{}_{k}F_{m}-2F^{m}P_{m\,k}\right\}
  =
  d\omega_{k}\,,
\end{equation}

\noindent
and the Komar charge is given by

\begin{equation}
  \label{eq:Komarcharge}
    \mathbf{Q}[k] -\omega_{k}
        =
        \star (e^{a}\wedge e^{b})
        e^{-2\phi}P_{k\, ab} -2\left(P^{m}{}_{k}F_{m}-P_{m\,k}F^{m}\right)\,.
\end{equation}

Observe that the electromagnetic terms occur in a symplectic-invariant
combination now. This hints at the electric-magnetic (SL$(2,\mathbb{R})$)
invariance of the Komar charge, a fact that we will study in
Section~\ref{sec-sduality}. Before studying this invariance, we are going to
check the validity of this formula in the family of static black holes
introduced in Section~\ref{sec-staticBHs} by direct computation of the Komar
integral.

\section{Checking the Smarr formula for static axion-dilaton black holes}
\label{sec-checking}

Now we want to compute the Komar integrals over the bifurcation sphere on the
horizon and over a sphere at spatial infinity for the static axion-dilaton
black holes introduced in Section~\ref{sec-staticBHs}. Thus, we are interested
in the $\theta\varphi$ components of the integrand only.  We compute them term
by term and we recover the normalization factor $(16\pi)^{-1}$. First,

\begin{equation}
    \star (e^{a}\wedge e^{b})P_{k\,ab}
     =
    \frac{1}{2\sqrt{|g|}}\varepsilon_{\mu\nu\rho\sigma}\nabla^{\mu}k^{\nu}
    dx^{\rho}\wedge dx^{\sigma}\,,
\end{equation}

\noindent
and, for these solutions

\begin{subequations}
  \begin{align}
\nabla^{\mu}k^{\nu}
& =
\delta^{t[\mu}\delta^{\nu]r}\partial_{r}e^{2U}\,,  
    \\
    & \nonumber \\
   \star (e^{a}\wedge e^{b})
    P_{k\,ab}
    & =
    -r^{2}\partial_{r}e^{2U}
    \sin{\theta} d\theta \wedge d\varphi\,.
\end{align}
\end{subequations}

The electric and magnetic momentum maps can be taken to be

\begin{equation}
  P^{m}{}_{k}=A^{m}{}_{t}\,,
  \hspace{1cm}
  P_{m\,k}=A_{m\, t}\,,
\end{equation}

\noindent
and, the second term in the Komar charge Eq.~(\ref{eq:Komarcharge}) is
(only $\theta\varphi$ components)

\begin{equation}
  \begin{aligned}
    -2\left(P^{m}{}_{k}F_{m\, \theta\varphi}-P_{m\,k}F^{m}{}_{\theta\varphi}\right)
    & =
    -2\left\{A^{m}{}_{t}
      \left[e^{-2\phi}\left(\star
          F^{m}\right)_{\theta\varphi}
        +aF^{m}{}_{\theta\varphi}\right]-A_{m\,t}F^{m}{}_{\theta\varphi}\right\}
    \\
    & \\
    & =
\left\{    2R^{2}e^{-2\phi}A^{m}{}_{t}\partial_{r}A^{m}{}_{t}
  +4e^{\phi_{\infty}}\left(A_{m\,t}-aA^{m}{}_{t}\right)\Im\mathrm{m}(\Gamma^{m})
  \right\}\sin{\theta}\,.
  \end{aligned}
\end{equation}

Integrating over a 2-sphere of constant radius $r$, we get

\begin{equation}
  \label{eq:Komarintegral-r}
  \mathcal{K}(S^{2}{}_{r})
  =
  \tfrac{1}{4}r^{2}\partial_{r}e^{2U}
  -\tfrac{1}{2}R^{2}e^{-2\phi}A^{m}{}_{t}\partial_{r}A^{m}{}_{t}
  -e^{\phi_{\infty}}\left(A_{m\,t}-aA^{m}{}_{t}\right)\Im\mathrm{m}(\Gamma^{m})\,.
\end{equation}

At infinity, only the first term contributes, giving

\begin{equation}
  \mathcal{K}(S^{2}{}_{\infty})
  =
  M/2\,.
\end{equation}

\noindent
Over the bifurcation sphere\footnote{Actually, it is enough to set $r=r_{+}$},
the first term gives $ST=r_{0}/2$, but we have to evaluate carefully the
second and third terms. We introduce some notation:

\begin{equation}
  A \equiv \lambda_{\infty}r +\lambda_{\infty}^{*}\Upsilon\,,
  \hspace{1cm}
  B \equiv r +\Upsilon\,,
  \,\,\,\,\,
  \Rightarrow
  \,\,\,\,\,
  \lambda=A/B\,.
\end{equation}

The second term in Eq.~(\ref{eq:Komarintegral-r}) is

\begin{equation}
  \begin{aligned}
    -\tfrac{1}{2}R^{2}e^{-2\phi}A^{m}{}_{t}\partial_{r}A^{m}{}_{t}
    & =
    \frac{1}{2R^{2}|r+\Upsilon|^{2}}\left[\Gamma^{m}B+\mathrm{c.c}\right]
    \left[-2|r+\Upsilon|^{2}\Re\mathrm{e}(\Gamma^{m})
      +4\Im\mathrm{m}(\Gamma^{m})\Im\mathrm{m}(\Upsilon)r\right]
    \\
    & \\
    & =
    \frac{1}{2R^{2}}\left[\Gamma^{m}(\Gamma^{m}+\Gamma^{m\,*})B+\mathrm{c.c}\right]
    \\
    & \\
    & \hspace{.5cm}
    +\frac{r}{R^{2}|r+\Upsilon|^{2}}\left[i\Gamma^{m}(\Gamma^{m}-\Gamma^{m\,*})B+\mathrm{c.c}\right]
\Im\mathrm{m}(\Upsilon)\,.
  \end{aligned}
\end{equation}

\noindent
Using the relation Eq.~(\ref{eq:upsilongamma}) it is not hard to see that at
$r=r_{+}$

\begin{equation}
i\Gamma^{m}(\Gamma^{m}-\Gamma^{m\,*})B(r_{+})+\mathrm{c.c}
  =
  -\tfrac{1}{2}R^{2}(r_{+})\Im\mathrm{m}(\Upsilon)\,.
\end{equation}

\noindent
Then,

\begin{equation}
  \begin{aligned}
\left.    -\tfrac{1}{2}R^{2}e^{-2\phi}A^{m}{}_{t}\partial_{r}A^{m}{}_{t}\right|_{r_{+}}
    & =
    \frac{2|\Gamma|^{2}r_{+}-M|\Upsilon|^{2}}{2R^{2}(r_{+})}
    -\frac{(Mr_{+}-2|\Gamma|^{2})}{2R^{2}(r_{+})}\Re\mathrm{e}(\Upsilon)
    -\frac{r_{+}[\Im\mathrm{m}(\Upsilon)]^{2}}{2|r_{+}+\Upsilon|^{2}}
    \\
    & \\
    & =
    \frac{r_{-}}{4} +\frac{\Re\mathrm{e}(\Upsilon)}{4}
    -\frac{r_{+}[\Im\mathrm{m}(\Upsilon)]^{2}}{2|r_{+}+\Upsilon|^{2}}\,.
  \end{aligned}
\end{equation}

The third term in  Eq.~(\ref{eq:Komarintegral-r}) is

\begin{equation}
  \begin{aligned}
-e^{\phi_{\infty}}\left(A_{m\,t}-aA^{m}{}_{t}\right)\Im\mathrm{m}(\Gamma^{m})
& =
-\frac{e^{2\phi_{\infty}}}{R^{2}}\left[\Gamma^{m}(A-aB) +\mathrm{c.c.} \right]
\Im\mathrm{m}(\Gamma^{m})
\\
& \\
& =
-\frac{e^{-2(\phi-\phi_{\infty})}}{R^{2}}\left[i\Gamma^{m}B +\mathrm{c.c.} \right]
\Im\mathrm{m}(\Gamma^{m})
\\
& \\
& =
\frac{1}{2|r+\Upsilon|^{2}}\left[\left(M\Re\mathrm{e}(\Upsilon)
    +2|\Gamma|^{2}\right)r +M|\Upsilon|^{2}+2|\Gamma|^{2}\Re\mathrm{e}(\Upsilon) \right]\,.
\end{aligned}
\end{equation}

Combining these two partial results at $r=r_{+}$ and operating, we get

\begin{equation}
  \begin{aligned}
&\,\,\,\,    \frac{r_{-}}{4} -\frac{\Re\mathrm{e}(\Upsilon)}{4}
+\frac{\left(M\Re\mathrm{e}(\Upsilon)
    +2|\Gamma|^{2}-[\Im\mathrm{m}(\Upsilon)]^{2}\right)r_{+}
  +M|\Upsilon|^{2}+2|\Gamma|^{2}\Re\mathrm{e}(\Upsilon)}{2|r_{+}+\Upsilon|^{2}}
\\
& \\
& =
  \frac{r_{-}}{4} 
  +\frac{\Re\mathrm{e}(\Upsilon)\left(2Mr_{+}+4|\Gamma|^{2}
      -|r_{+}+\Upsilon|^{2}\right)
    +2\left(2|\Gamma|^{2}-[\Im\mathrm{m}(\Upsilon)]^{2} \right)r_{+}
  +2M|\Upsilon|^{2}}{4|r_{+}+\Upsilon|^{2}}
\\
& \\
& =
  \frac{r_{-}}{4} 
  +\frac{\Re\mathrm{e}(\Upsilon)\left(2Mr_{+}-r_{+}^{2}+4|\Gamma|^{2}
      -|\Upsilon|^{2}\right)
    +2\left(2|\Gamma|^{2}-|\Upsilon)|^{2} \right)r_{+}
  +2M|\Upsilon|^{2}}{4|r_{+}+\Upsilon|^{2}}
\\
& \\
& =
  \frac{r_{-}}{4} 
  +\frac{2\Re\mathrm{e}(\Upsilon)r_{+}r_{-}
      +r_{+}^{2}r_{-} -|\Upsilon)|^{2}r_{-}}{4|r_{+}+\Upsilon|^{2}}
\\
& \\
& =
  \frac{r_{-}}{2}\,, 
\end{aligned}
\end{equation}

\noindent
which gives the Smarr formula proposed in Section~\ref{sec-staticBHs},
Eqs.~(\ref{eq:Smarrformula}).

\section{Charges, potentials and S~duality}
\label{sec-sduality}

The static axion-dilaton black holes introduced in Section~\ref{sec-staticBHs}
are the most general black holes in that class according to the no-hair
theorems because they have the maximum number of independent parameters
(moduli $\lambda_{\infty}$ and conserved charges $M,\Gamma^{m}$) allowed by
it. Hence, we have proven the validity of the Smarr formula in this theory for
static black holes. However, we have to rewrite it in terms of the potentials
and charges.

The charges which are quantized in this theory are not the components
of $\Gamma^{m}$, but

\begin{subequations}
  \begin{align}
    p^{m}
    & \equiv 
      \frac{1}{8\pi G_{N}^{(4)}}\int F^{m}
      =
      e^{\phi_{\infty}}\Im\mathrm{m}(\Gamma^{m})/G_{N}^{(4)}\,,
    \\
    & \nonumber \\
    q_{m}
    & \equiv 
      \frac{1}{8\pi G_{N}^{(4)}}\int F_{m}
      =
      e^{\phi_{\infty}}\left[e^{-2\phi_{\infty}}\Re\mathrm{e}(\Gamma^{m})
    +a_{\infty}\Im\mathrm{m}(\Gamma^{m})\right]/G_{N}^{(4)}\,.
  \end{align}
\end{subequations}

\noindent
According to the discussions in
Refs.~\cite{Elgood:2020svt,Elgood:2020mdx,Elgood:2020nls}, the potentials can
be identified, up to a normalization factor, with the momentum maps
$P^{m}{}_{k}$ and $P_{m\,k}$ evaluated over the black-hole horizon:

\begin{subequations}
  \begin{align}
\Phi^{m}
    & \equiv 
    2\left.P^{m}{}_{k}\right|_{r_{h}}\,,
    \\
    & \nonumber \\
\Phi_{m}
    & \equiv 
    2\left.P_{m\,k}\right|_{r_{h}}\,,
  \end{align}
\end{subequations}

\noindent
and they are guaranteed to be constant at least over the bifurcation sphere
$\mathcal{BH}$, according to the \textit{restricted, generalized zeroth
  laws}.\footnote{This result may be extended to the complete event horizon
  using the arguments in Ref.~ \cite{Jacobson:1993vj}.}  We normalize them to
vanish at infinity for the asymptotically-flat solutions we are interested in.

Therefore,

\begin{subequations}
  \begin{align}
    \frac{1}{16\pi G_{N}^{(4)}}
    \int_{S^{2}_{\infty}}2\left(P^{m}{}_{k}F_{m}-P_{m\,k}F^{m}\right)
    & =
      0\,,
    \\
    & \nonumber \\
    \frac{1}{16\pi G_{N}^{(4)}}
    \int_{\mathcal{BH}}2\left(P^{m}{}_{k}F_{m}-P_{m\,k}F^{m}\right)
    & =
     \tfrac{1}{2}\left( \Phi^{m}q_{m}-\Phi_{m}p^{m}\right)\,.
  \end{align}
\end{subequations}

\noindent
On the other hand, on general grounds and in the static case,

\begin{subequations}
  \begin{align}
    -\frac{1}{16\pi G_{N}^{(4)}}
    \int_{S^{2}_{\infty}} \star (e^{a}\wedge e^{b}) e^{-2\phi}P_{k\, ab}
    & =
      M/2\,,
    \\
    & \nonumber \\
    -\frac{1}{16\pi G_{N}^{(4)}}
    \int_{\mathcal{BH}}\star (e^{a}\wedge e^{b}) e^{-2\phi}P_{k\, ab}
    & =
      ST\,,
  \end{align}
\end{subequations}

\noindent
and the Smarr formula takes the general form\footnote{A previous derivation of
  a Smarr formula in this theory was made in Ref.~\cite{Rogatko:1994tv} and
  our results should be compared with those in that reference.}

\begin{equation}
  \label{eq:generalSmarr}
M= 2ST +\Phi^{m}q_{m}-\Phi_{m}p^{m}\,.   
\end{equation}

While our definitions of charges and potentials seem to be identical to those
in Refs.~\cite{Breitenlohner:1987dg,Gibbons:1996af}, we get a different sign
for the last term. The scaling arguments explained in
Refs.~\cite{Kastor:2008xb,Kastor:2010gq} indicate that the sign should be a
plus if we define $\Phi^{m}=\partial M/\partial q_{m}$. We can always add a
sign to our definition of $\Phi_{m}$ to make it coincide with that definition,
but we are going to argue that a relative minus sign between the last two
terms is the natural sign if we take into account that the Smarr formula
should be invariant under the dualities of the theory. These always act on the
vector fields of a 4-dimensional theory through a symplectic embedding
\cite{Gaillard:1981rj}.

In this particular case, it is convenient to define the symplectic vector of
field strengths as follows:

\begin{equation}
  \left(\mathcal{F}^{M}\right)
  \equiv
  \left(
    \begin{array}{c}
      F_{m} \\ F^{m} \\
    \end{array}
  \right)\,,  
\end{equation}

\noindent
since the action of a SL$(2,\mathbb{R})\sim$Sp$(2,\mathbb{R})$ duality
transformation

\begin{equation}
S
\equiv
\left(
 S^{M}{}_{N}
  \right)
=
\left(
  \begin{array}{cc}
    \alpha  & \beta \\ \gamma & \delta \\
  \end{array}
  \right)\,,
\end{equation}

\noindent
on them and on the axidilaton takes a simpler form:

\begin{equation}
  \mathcal{F}^{\prime\, M}
  =
  S^{M}{}_{N}\mathcal{F}^{N}\,,
  \hspace{1cm}
  \lambda'
  =
  \frac{\alpha\lambda+\beta}{\gamma\lambda+\delta}\,,
  \hspace{1cm}
  \alpha\delta-\beta\gamma=1\,.
\end{equation}

It follows from the definitions that 

\begin{equation}
  \left(\mathcal{P}^{M}{}_{k}\right)
  \equiv
  \left(
    \begin{array}{c}
 P_{m\, k} \\    P^{m}{}_{k} \\
    \end{array}
    \right)\,,
  \hspace{.7cm}
  \left(\Phi^{M}\right)
  \equiv
  \left(
    \begin{array}{c}
 \Phi_{m} \\    \Phi^{m} \\
    \end{array}
    \right)\,,
  \hspace{.7cm}
  \left(\mathcal{Q}^{M}\right)
  \equiv
  \left(
    \begin{array}{c}
  q_{m} \\ p^{m} \\
    \end{array}
    \right)\,,
\end{equation}

\noindent
transform as the  SL$(2,\mathbb{R})$ vector $\mathcal{F}^{M}$. 

An important property of the duality group SL$(2,\mathbb{R})$ is that
it is isomorphic to Sp$(2,\mathbb{R})$ since the condition

\begin{equation}
  S^{M}{}_{P}\Omega_{MN}S^{N}{}_{Q}= \Omega_{PQ}\,,
  \hspace{1cm}
\left(
 \Omega_{MN}
  \right)
=
\left(
  \begin{array}{cc}
    0  & 1 \\ -1 & 0 \\
  \end{array}
  \right)\,,
\end{equation}

\noindent
also implies $\alpha\delta-\beta\gamma=1$ for the matrix $S$. Thus,
the combination of potentials and charges occurring in the Smarr
formula Eq.~(\ref{eq:generalSmarr})

\begin{equation}
  \Phi^{m}q_{m}-\Phi_{m}p^{m}= \mathcal{Q}^{M}\Phi^{N}\Omega_{MN}
\end{equation}

\noindent
is manifestly SL$(2,\mathbb{R})\sim$Sp$(2,\mathbb{R})$-invariant.  The
explicit calculation of this term in Section~\ref{sec-checking} is a
proof of this invariance.

\section{Discussion}
\label{sec-discussion}

In this paper we have shown how the momentum maps introduced in
Ref.~\cite{Elgood:2020svt,Elgood:2020mdx,Elgood:2020nls} in the context of
black-hole thermodynamics can be used to express the Komar integral obtained
in the context of Wald's formalism \cite{Liberati:2015xcp} as a surface
integral in a manifestly covariant way, generalizing the results of
\cite{Kastor:2008xb,Kastor:2010gq,Ortin:2021ade}. We have also shown how, in
its turn, this integral can be used to derive a Smarr formula which is
manifestly symplectic invariant. We have checked this formula explicitly in
the most general family of static axidilaton black holes, constructed in
Ref.~\cite{Kallosh:1993yg}. It is trivial to extend these results to theories
with more scalars and more complicated kinetic matrices (period matrices in
the language of $\mathcal{N}=2$ theories).

Symplectic invariance is a property to be expected of a general Smarr formula
because this relation is just a relation between physical parameters occurring
in the metric, which is symplectic invariant. It is, nevertheless, surprising,
how this property of the Smarr formula and of the Komar integral from which it
is derived, arises from a combination of the Noether-Wald charge and the
on-shell Lagrangian density which are not separately symplectic invariant. The
lack of symplectic invariance of the Noether-Wald charge leads to a
non-invariant first law Ref.~\cite{Elgood:2020svt} in which the terms
containing the variations of the magnetic charges are not present. Since
Wald's formalism is based on gauge symmetries and there is no gauge symmetry
associated to the conservation of magnetic charges (at least in the standard,
off-shell, formulation of electromagnetism and its generalizations) this was
to be expected. It is, nevertheless, somewhat unsatisfactory.\footnote{It is
  also problematic, since this is the only formalism that can be applied to
  theories of higher order in the curvature.} It is also somewhat
unsatisfactory that there is no natural place in Wald's first law for terms
proportional to the variations of the asymptotic values of scalars
\cite{Gibbons:1996af} since these are not associated to gauge symmetries
either. Work on these problems is in progress.

\section*{Acknowledgments}

This work has been supported in part by the MCIU, AEI, FEDER (UE) grant
PGC2018-095205-B-I00 and by the Spanish Research Agency (Agencia Estatal de
Investigaci\'on) through the grant IFT Centro de Excelencia Severo Ochoa
SEV-2016-0597. The work of DP is supported by a ``Campus de Excelencia
Internacional UAM/CSIC'' FPI pre-doctoral grant. TO wishes to thank
M.M.~Fern\'andez for her permanent support.


\end{document}